\documentclass[useAMS,usenatbib]{mn2e}

\usepackage{graphicx}
\usepackage{amssymb}
\usepackage{longtable}
\usepackage{ulem}
\usepackage{color}

\newcommand{\civ}{C~IV}
\newcommand{\lya}{Ly$\alpha$}
\newcommand{\Hb}{H$\beta$}
\newcommand{\ox}{$\alpha_{\rm ox}$}
\newcommand{\mox}{\alpha_{\rm ox}}
\newcommand{\LL}{$L_{\rm Bol}/L_{\rm Edd}$}
\newcommand{\mLL}{L_{\rm Bol}/L_{\rm Edd}}

\newcommand{\Lion}{$L_{\rm ionise}$}

\newcommand{\mLline}{L({\rm line})}
\newcommand{\mLion}{L_{\rm ionise}}

\newcommand{\covfac}{$\Omega/4\pi$}
\newcommand{\mbh}{$M_{\rm BH}$}

\title[The environment of WLQs]{The environment of Weak Emission-Line Quasars}

\author[M. Niko\l ajuk and R. Walter]{M. Niko\l ajuk$^{1,2}$\thanks{E-mail:
Marek.Nikolajuk@unige.ch, mrk@alpha.uwb.edu.pl (MN); Roland.Walter@unige.ch (RW)} and R.
Walter$^{1}$\footnotemark[1]
\\
$^{1}$ISDC Data Centre for Astrophysics, Chemin d'Ecogia 16,
                CH-1290 Versoix, Switzerland \\
$^{2}$Faculty of Physics, University of Bialystok, Lipowa
                41, PL-15424 Bia\l ystok, Poland}
\begin{document}

\date{Accepted 2011 November 16; in original form 2011 July 26}

\pagerange{\pageref{firstpage}--\pageref{lastpage}} \pubyear{2011}

\maketitle

\label{firstpage}
\begin{abstract}

The nature of weak emission-line quasars (WLQs) is probed by 
comparing the Baldwin effect (BEff) in WLQs and normal quasars (QSOs).
We selected 81 high-redshift (z $>$ 2.2) and 2 intermediate-redshift 
(z = 1.66 and 1.89) WLQs. Their rest-frame equivalent widths (EWs) of
the \civ\ emission-line and their Eddington ratio were obtained from 
the Sloan Digital Sky Survey Data Release 7 (SDSS DR7) Quasar Catalogue
or from \citeauthor{DS2009}. We compare the parameters of WLQs with these 
of 81~normal quasars from Bright Quasar Survey (BQS) and 155~radio-quiet 
and radio-intermediate quasars detected by SDSS and Chandra. 
The influence of the Eddington ratio, \LL, and the X-ray to optical 
luminosity ratio, \ox, on the BEff is analysed. 
We find that WLQs follow a different relationship on the EW(\civ)-\LL\ 
plane than normal quasars. This relationship disagrees with the 
super-Eddington hypothesis. The weakness/absence of emission-lines 
in WLQs does not seem to be caused by their extremely soft ionizing 
continuum but by low covering factor (\covfac) of their broad line 
region (BLR). Comparing emission-line intensities indicates that 
the ratios of high-ionization line and low-ionization line
regions (i.e. $\Omega_{\rm HIL}/\Omega_{\rm LIL}$) are lower in WLQs
than in normal QSOs. The covering factor of the regions producing 
\civ\ and \lya\ emission-lines are similar in both WLQs and QSOs.
\end{abstract}


\section{Introduction}

A negative correlation between the broad emission-line 
equivalent width (EW) and the luminosity in active galactic nuclei (AGNs) 
was discovered by \citet{Baldwin77} for the \civ\ $\lambda1549$ line.
Similar correlations (the Baldwin effect, hereafter BEff) are also observed
for other lines such as \lya$\lambda$1216, Si~IV+O IV $\lambda$1400, 
He~II $\lambda$1640,4686, C~III] $\lambda$1909, Mg~II $\lambda$2800, 
Fe lines in the UV and optical bands, and the Balmer lines produced 
in the broad line region (BLR)
\citep[see e.g.][]{Kinney90,Zamorani92,Green2001,Kuraszkiewicz2002,Shang2003}. 
This effect was also observed in single objects (e.g. NGC 5548, NGC4151), 
when the intrinsic ionizing continuum is varying
\citep{Kinney90,PP92,GP2003,Kong2006}. At least some of the emission-lines 
produced in the narrow line region display a BEff as well 
\citep[e.g.][]{BG92,Dietrich2002,Kereme2009}. 
Furthermore, an X-ray BEff in the iron K lines was detected by \citet{IT93} 
and analysed by e.g. \citet{Jiang2006,Bianchi2007}.

Several physical explanations have been proposed to explain the BEff. 
The most supported hypothesis is that the more luminous objects
have softer UV/X-ray spectra reducing ionization and photoelectric 
heating in the BLR gas. Ipso facto the equivalent widths (EWs) of 
emission-lines are reduced at higher luminosity with the strongest 
effect for high-ionization lines (HILs) (see \citealt{KBF98} and 
\citealt{Shields2007} for a review of the BEff).
Fundamental parameters such as the Eddington ratio 
\citep{BL2004,Warner2004,Bachev2004,ZW2005,Xu2008,Dong2009}, 
the black hole mass \citep{Netzer92,Shields2007}, or metalicity 
\citep{Warner2004} have also been suggested as the driver of the BEff. 

The discovery of weak emission-line quasars (WLQs) 
i.e. sources with abnormally low broad emission-lines
\citep[e.g. EW(\lya)$_{\rm WLQ} < 15.4$ \AA,][]{DS2009} provides
new constraints on the driving mechanism of the BEff. 
The first WLQ object -- PG~1407+265 (with redshift z $=0.94$) 
was discovered by \citet{McDowell95}. 
However, up to 2009 only about 20 WLQs were known.
They mostly lie at high-redshifts ($ \mathrm{z} > 2.2$), like 
SDSS~J153259.96-003944.1 \citep[][the first high-z WLQ, with
$\mathrm{z} = 4.62$]{Fan99} and were found in the Sloan Digital Sky 
Survey (SDSS) \citep{Anderson2001,Schneider2003,Schneider2005,Collinge2005,
Fan2006,Schneider2007,Shemmer2009}. \citet{DS2009} recently discovered 
65 new high-z WLQs, which may suggest that there is a deficit of weak line 
quasars below z $< 2$. However, \citet{Plotkin2010a,Plotkin2010b} 
pointed out that more intermediate- and low-redshift WLQ may also exist.

There is no generally accepted explanations for the weakness or even 
absence of emission-lines in WLQ. Several hypothesis were suggested 
by \citet{McDowell95}. Relativistic beaming in WLQ is not favoured as 
weak line quasars, in contrast to BL Lacs, are radio-quiet 
objects\footnote{The radio-loudness parameter R is defined as 
the ratio of the rest-frame 6 cm to 2500 \AA\ flux densities 
\citep[see][]{Jiang2007,Shen2011}. Among 70 radio detected WLQs 
analysed by \citet{DS2009} there is 81\% of sources with 
$\mathrm{R} \le 25$ and only 7\% of radio-loud WLQs i.e. with 
$\mathrm{R} > 100$.}, show no variability or strong polarization. 
Moreover, the radio spectral slopes, connecting $\lambda \sim$6 cm with 
$\sim$20 cm, are significantly steeper in radio-detected WLQs than the 
typical slopes for BL Lac, $\alpha_{\rm r} \sim 0.3$ 
\citep{Shemmer2006,Shemmer2009,DS2009,Plotkin2010a}.
The idea, that WLQs could be broad absorption line (BAL) quasars also 
meet difficulties. Generally, they do not show broad absorption features 
and are clasiffied as non-BAL objects \citep{DS2009,Shen2011}.

Two leading hypothesis to explain the weakness of emission-lines 
in WLQ have been suggested: 

(1) The first one is related to the BEff. 
Weak emission-lines may be a consequence of a very soft ionizing continuum 
and of a relative deficiency of high-energy UV/X-ray photons
\footnote{``the very soft ionizing continuum'' means that  
a continuum in the far-UV (FUV) band is characterized by a steep spectrum.
We use the X-ray to optical luminosity ratio, \ox (see definition below).
For typical quasar \ox\ is equal to $-1.50$ \citep{Laor97} and for the
most luminous quasars with redshift 1.5-4.5 the mean \ox\ equals 
to -1.80 \citep{Just2007}. We adopt $\mox < -2.0$ as a definition 
of a very soft spectrum.}.
\citet{Leighly2007a}, based on observation of PHL 1811, have claimed 
that its very soft spectral energy distribution (SED) (the photon index 
$\mox = 2.3 \pm 0.1$)
\footnote{
$\mox = 0.3838 \log[L_{\nu}(2 \mathrm{keV})/L_{\nu}(2500 \mathrm{\AA})]$
\citep[e.g.][]{AT82,Strateva2005,Gibson2008}.}
is related to its super-Eddington nature (the estimated \LL\
lie in the range 0.9-1.6). However, it is worth noting that, 
the observed UV/optical part of the continuum in WLQs looks like
these of normal quasars. \citet{Richards2003} have analysed the
spectra of 4576 SDSS quasars and they found that the 
spectral indices, $\alpha_{\nu}$ (where $f_{\nu} \propto \nu^{\alpha_{\nu}}$), 
lie in a wide range, with mean values from $-0.25$ to $-0.76$
(see their composites no. 1-4). The spectral indices in WLQs 
also span the same interval with a median values of $\alpha_{\nu} = -0.52$
\citep{DS2009,Plotkin2010a}. Those values means that, generally, 
the observed UV SED in WLQ is not particularly soft. However, these objects 
may still emit more vigorously in the unobserved far-UV band. 
Recently \citet{Wu2011} found a small population of X-ray weak quasars,
suggesting that these PHL 1811 analogs possess the shielding gas with 
large covering factor. This gas absorbs almost all soft X-ray continuum 
to prevent illumination of broad line region (BLR) by this radiation. 
As a result weak emission-lines are produced although a face-on observer 
see the normal X-ray continuum.

(2) The second hypothesis suggests that WLQs are normal quasars with 
typical  metallicities, ionizing continua, and ionization parameters,
however, with an underdeveloped BLR perhaps because of 
a freshly launched accretion disc wind \citep{Hryn2010}. 
The weakness/absence of emission-lines in this case is caused by 
a low BLR covering factor or a deficit of line-emitting gas in the BLR 
\citep{Shemmer2010}.

In this paper we analyse both hypothesis: softness of ionizing 
continuum and underdevelopment of BLR. In section~\ref{sec:data} 
we describe the sample of quasars that was used. 
Section~\ref{sec:res} is devoted to the comparison of the observed 
properties of WLQ and QSO that we discuss in section~\ref{sec:dis}.
The conclusions are presented in section~\ref{sec:con}.
We assume that $H_0 = 70$ km s$^{-1}$ Mpc$^{-1}$, $\Omega_m = 0.3$,
and $\Omega_{\Lambda} = 0.7$.

\section{The WLQ sample}
\label{sec:data}

The sample of WLQs consists of 81 high-redshift (z $> 2.2$)  
and high-bolometric luminosity $(\log L_{\rm Bol} = 46.5-47.9)$
sources classified by 
\citet{Anderson2001,Collinge2005,Schneider2007,Shemmer2009,DS2009,Plotkin2010a},
extended by two quasars
SDSS J094533.98+100950.1 (hereafter SDSS J094534) 
and SDSS J170108.89+395443.0 (hereafter SDSS J170109).
Both sources lie at intermediate-redshift distances with z of 
1.66 and 1.89, respectively.

\begin{figure}
\centering
\includegraphics[width=0.5\textwidth]{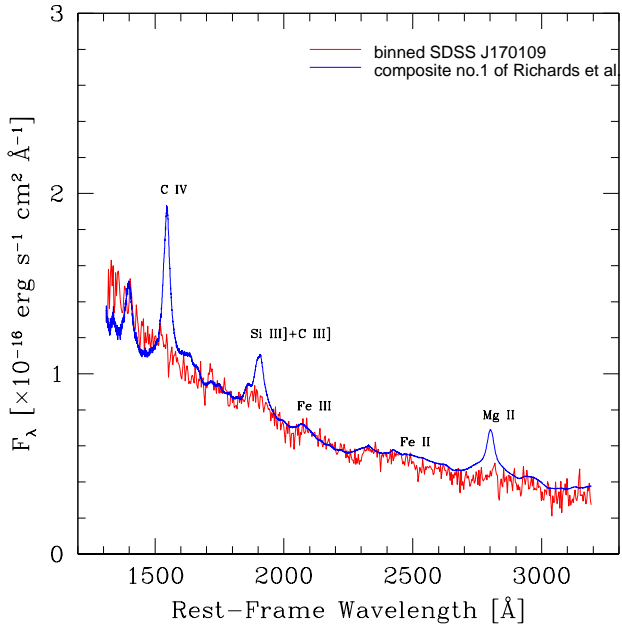}
\caption{The rest-frame spectrum of SDSS J170108.89+395443.0 binned 
and corrected for Galactic reddening using \citet{Cardelli89} relationship. 
For comparison, \citet{Richards2003} composite spectrum (no.~1) is shown.}
\label{fig:J170109}
\end{figure}

A spectrum of SDSS J094534 was analysed extensively by \citet{Hryn2010}. 
The second source (see its spectrum in Fig.~\ref{fig:J170109})
was retrieved serendipitously by us in the SDSS Data Release~7 (DR7)
quasar catalogue \citep{Shen2011}. We classified it as WLQ 
because (1) the equivalent widths of the \civ\ and 
Mg II measured at the rest frame are small (i.e. EW(\civ) 
$= 2.09 \pm 1.83$\AA, EW(Mg II) $= 9.41 \pm 2.03$\AA, 
\citealt{Shen2011}), (2) this quasar is radio-intermediate as
a dozen WLQ sources (rest-frame $f_{\rm 6cm}/f_{2500} = 45.3$, 
\citealt{Shen2011}), (3) the UV continuum of SDSS~J170109 can be fit 
as a power law $(f_{\nu} \propto \nu^{\alpha_{\nu}})$ with a spectral index  
$\alpha_{\nu} = -0.23 \pm 0.03$\footnote{This value is equivalent to 
$\alpha_{\lambda} = -1.77$ where $f_{\lambda} \propto \lambda^{\alpha_{\lambda}}$.}
identical within errors to that of the quasar composite from 
\cite{Richards2003}. This value also 
differs from the mean spectral index calculated for BL Lac candidates 
for which we have $\langle \alpha_{\nu} \rangle = -1.15$ \citep{Plotkin2010a}.

All sources were detected by the Sloan Digital Sky Survey.
In our analysis we are using the equivalent widths, EW, of  
emission-lines measured in the rest-frame, fluxes of those lines, 
masses of the supermasive black holes, $M_{\rm BH}$, 
accretion rates in the Eddington units, \LL, and the spectral 
indices, \ox. Almost all but \ox\ values, were found in 
the SDSS~DR7 quasar catalogue \citep{Shen2011}. 
\citet{Shen2011} point out that estimated 
EW(\civ) are encumbered with large error when the signal-to-noise 
ratio (S/N) of the observed WLQ spectrum is lower than 5 (see their figure~8). 
Therefore, in these cases we use EW values estimated by \citet{DS2009} 
which for all sources but two have EW(\civ) $> 5 \sigma$.
In other cases we use upper limits taken from the quasar 
catalogue or calculate them (see Table~\ref{tab:WLQ}). 
The spectral indices, \ox, of WLQs were taken from 
\citet{Shemmer2006,Shemmer2009}. 
All those values originate from the Chandra observations.
Additionally, we checked the Chandra Multiwavelength Project 
Catalogue \citep{Green2009}. We cross-correlated this catalogue 
with SDSS DR7. No new WLQs but SDSS~J170109
were found. Its \ox\ is equal to $-1.29$.

\section{Comparison of WLQs with normal quasars}
\label{sec:res}

\begin{figure}
\centering
\includegraphics[width=0.5\textwidth]{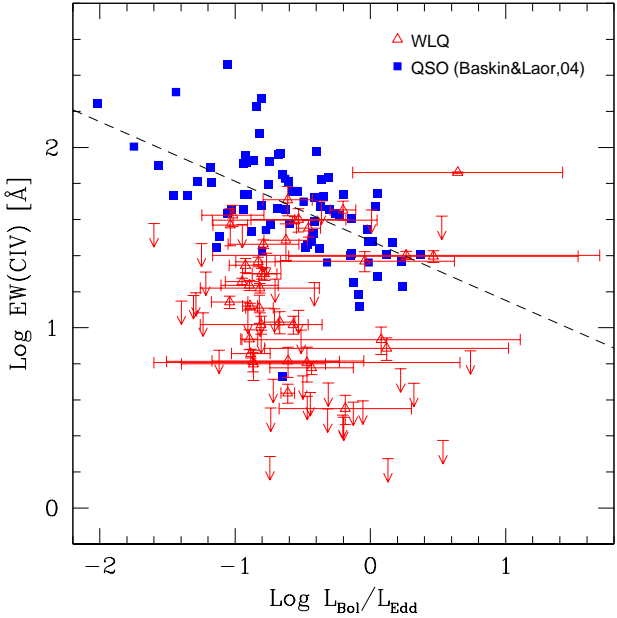}
\caption{Equivalent width of C IV measured at the rest frame
plotted against accretion ratio. Filled blue squares show 81 BQS 
quasars analysed by \citet{BL2004}. Filled red triangles and 
upper limits refer to 76~WLQs taken from \citet{Shen2011} or \citet{DS2009}.
Points with error bars refer to objects with a significance of EW
higher than 5$\sigma$. Otherwise, upper limits are shown.
Dashed line is the best linear fit to BQS quasars 
\citep{BL2004,BL2005}.
}
\label{fig:mC4all}
\end{figure}

Our aim is to compare the Baldwin effect observed in weak emission-line 
quasars to that observed in normal Type 1 quasars. Fig.~\ref{fig:mC4all} 
displays the EW of the \civ\ emission-line against the dimensionless 
accretion rates for different types of quasars.
This figure includes 81 quasars from the Bright Quasar Survey (BQS) 
with redshifts z $< 0.5$ and bolometric values $\log L_{\rm Bol} = 
44.2-46.9$ analyzed by \citet{BG92}. \citet{BL2004} estimated their 
EW(\civ) and \LL, respectively. Dashed line represents the best 
linear fit to their data \citep{BL2004,BL2005}.
The triangles show 76 WLQs for which the EW and the accretion ratios 
were calculated by \citet{Shen2011} or \citet{DS2009}.
We must notice here that in both Baskin \& Laor's and 
Shen et al.'s papers the methods to estimate \LL\ are similar. 
Both calculated the bolometric luminosity using relationship 
$L_{\rm Bol} = \mathrm{BC_{\lambda}} \times L_{\lambda,\rm cont}$, where
$L_{\lambda, \rm cont}$ is the continuum luminosity measured at 
wavelength $\lambda$ and $\mathrm{BC_{\lambda}}$ is the appropriate 
bolometric correction factor. Both methods estimate the 
Eddington luminosity, $L_{\rm Edd} \propto M_{\rm BH}$ using the scalling 
method in order to calculate the black hole mass in AGN i.e. 
$M_{\rm BH} \propto  L_{\lambda, \rm cont}^{\rm b}$ FWHM$^2$(ion).
In this equation FWHM stands for the Full Width at Half Maximum of ion which 
produces the emission-line. BQS quasars and high-z WLQs lie at different 
distances, therefore, Baskin \& Laor and Shen et al. 
used observations of different emission-lines and continuum luminosities 
to calculate $M_{\rm BH}$. \citet{BL2004} used FWHM of H$\beta$ line, 
$L_{\lambda, \rm cont}$ measured at 5100 \AA\ in the rest-frame of quasar
and $\mathrm{b} = 0.50$ \citep[see equation (3) in ][]{Laor98}. 
The authors used H$\beta$ emission-line to estimate \LL\ because many 
scientists suggest non virialize character of \civ\ 
\citep[e.g.][]{Risaliti2009,Fine2010,Richards2011}.
However, high-z WLQs have redshifts higher 
than 2.2. Therefore, \citet{Shen2011} used \civ\ line and 
continuum luminosity observed at 1350\AA. They used the relationship
determined by \citet{VP2006} between $M_{\rm BH}$, 
FWHM, and $L_{\lambda, \rm cont}$ for which $\mathrm{b} = 0.53$.
\citet{BL2004} found an anti-correlation between EW(\civ) and \LL.
However, if one calculate Eddington ratios based on \civ\ emission-lines
this relationship is much weaker than the correlation 
with the \LL\ estimated based on \Hb\ \citep{BL2005}.

In this paper we analyse 83 weak emission-line quasars, however, 
in Fig.~\ref{fig:mC4all} only 76 of them have \civ\ emission-line 
strong enough to determine their \LL\ (see Table~\ref{tab:WLQ}). 
The EW of the remaining WLQs are lower than $\sim$3 to 7~\AA\ 
for sources with stronger or weaker UV fluxes, respectively.
It is worth noting that one object of the BQS lie in the region 
dominated by WLQ. This is the radio-quiet quasar (PG~0043+039, 
$z=0.384$) with EW(C IV) = $5.4 \pm 3.7$ \AA\ \citep{BL2004}.
Its H$\beta$ emission-line is strong with EW = 92 \AA, 
whereas, O~[III] $\lambda 5007$ and He~II $\lambda 4686$ 
are weak, with EWs equal to 1 and 0~\AA, respectively \citep{BG92}. 

On the contrary to PG~0043+039 several (17~from 76) WLQs 
have large equivalent widths of \civ\ line and behave like 
normal Type 1 quasars. Their median distance from 
the Baskin \&\ Laor's linear fit is only $2.3\sigma$ when 
for ``genuine'' WLQs is larger $(> 7\sigma)$.
We must mention, that \citet{Shemmer2009} decided to use EW(\civ) 
$\lesssim$ 10~\AA\ as a hallmark of WLQs. However, \citet{DS2009} 
decided to use EW of \lya+N~V blend as those lines are better seen in 
distant weak line quasars. Therefore, we kept this definition 
(i.e. EW(\lya+N~V) $< 15.4$~\AA) as a rule for our study.
We suggest that the weakness of the EW(\lya+N~V) in the sources 
with prominent \civ\ emission-lines is caused by strong absorption of 
the \lya\ region and that they are, in fact, normal quasars.

The best linear fit to BQS quasars sample (seen in Fig.~\ref{fig:mC4all})
suggests that WLQs follow a different relationship than normal quasars 
between EW and the accretion ratio. The errors of those quantities for 
WLQ are large and,unfortunately, we cannot fit a correlation to 
them. However, in order to statistically quantify the hypothesis 
about different relations we compare the reduced chi-squares, 
$\widetilde{\chi}^2$, in BQS' and WLQ's cases. We divide our WLQ 
objects into two subsets. The first one consists of all 76 weak 
emission-line quasars. In the second case we exclude all the upper 
limits on EW(\civ) from our subset. We also assume that that 
the obtained fit parameters for normal quasars are also the same 
for WLQs. The estimated reduced chi-square is 27.7 in the case 
of BQS quasars. That value is significantly larger than 1. 
However, we must notice that there is a large spread in distribution 
of normal quasars around the linear fit. If we assume that a natural 
spread is less than 9 \AA\ and we calculate the fit avoiding outliers
than the reduced chi-squares decreases to $\simeq 1.3$.  
The estimated $\widetilde{\chi}^2$ for WLQs are $\sim 1000$ including 
and $\sim 2100$ excluding upper limits on EW, respectively. 
The obtained values corroborate the hypothesis about difference 
in relationships. 

We must keep in mind, that \LL\ values are calculated using 
the method based on the luminosity-\mbh\ relation, i.e. 
$\mLL \propto \mathrm{FWHM(\civ)}^{-2}$. In many cases, 
the emission-lines in WLQ objects are broad and their FWHM
equals to a few thousand km s$^{-1}$ (see Mg~II in \citealt{Hryn2010},
\Hb\ in \citealt{Shemmer2010} or \civ\ in \citealt{Shen2011}).
Nevertheless, for weak \civ\ line (e.g. EW $<$ a few \AA) 
the FWHM value is underestimated thus \LL\ ratio is overestimated.

The existence of normal accretion rates in WLQs was discussed recently 
by \citet{Hryn2010}.  They have argued that when the Eddington ratio 
increases the width of \lya, Mg II lines decreases, \civ\ emission 
decreases, however, the Si IV line should become stronger, 
and the UV Fe II emission decreases. As the last two behaviours 
are not observed in WLQ 
\citep[see e.g.][]{Schneider2010}\footnote{\texttt{http://www.sdss.org/dr7/}}, 
\citet{Hryn2010} claimed that the weakness of the emission-lines 
in WLQs is not caused by high \LL.

\begin{figure}
\centering
\includegraphics[width=0.5\textwidth]{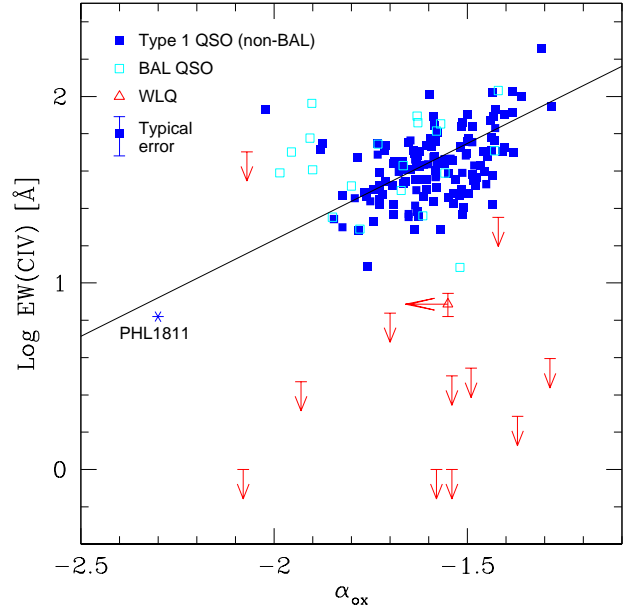}
\caption{Rest frame equivalent width of \civ\ emission-line versus 
spectral index \ox. Solid blue and open cyan squares refers to
Type 1 non-BAL and BAL quasars, respectively. Those 155 radio-quiet 
and radio-intermediate sources are taken from \citet{Green2009} paper. 
Solid red triangles and upper limits show 12 WLQ objects. 
Star denotes NLS1 PHL 1811. Solid line is the relationship 
obtained by \citet{Wu2009}. Typical error of non-BAL
and BAL QSOs is shown in the legend.
}
\label{fig:oxC4}
\end{figure}

So far, no observations of the FUV spectra of WLQs were made. 
Therefore, we analysed \ox\ which can shed light on the shapes of 
the SED in the FUV/soft X-ray band (Fig.~\ref{fig:oxC4}).
Apart from the spectral indices for WLQs we analyzed 
together with them 155 normal quasars.
Their \ox\ values were taken from the Chandra Multiwavelength 
Project \citep{Green2009}. We cross-corellated this catalogue with 
SDSS DR7 Quasar Catalogue \citep{Shen2011} to obtain the EW(\civ) 
of quasars. 
The solid line in Fig.~\ref{fig:oxC4} represents the best fit made 
by \citet{Wu2009}. We must notice that this linear fit was made to
another sample of quasars, however, it fits very well to our sample of 
normal quasars. Our study clearly shows that \ox\
values in weak emission-line quasars span the same region as seen in 
non-BAL and BAL QSOs. 
It points out that the UV/soft X-ray SED of WLQs is similar to 
those seen in normal AGNs and proves that a soft ionizing continuum
is not the reason for the weakness of the lines. That situation is 
found in PHL 1811 which is NLS1 galaxy with super-Eddington rate 
(\LL $\sim$ 0.9-1.6) and steep ionizing continuum ($\mox = -2.3$) 
\citep{Leighly2007b}. Therefore, PHL 1811 follows the relationship 
estimated by \citet{Wu2009}.

\section{Discussion}
\label{sec:dis}

The WLQs are shifted vertically in the $\log$ EW(\civ)-$\log$\LL\ plane 
relative to normal quasars (Fig.~\ref{fig:mC4all}).
This offset and the fact that QSOs and WLQs SED are almost the same, 
indicate that weak emission-line quasars are normal AGNs, however, 
with intrinsically weak \civ\ emission-line. 
It is also clearly shown that the super-Eddington luminosities are not 
required in weak line quasars contrasting with the idea that WLQs are 
super-Eddington sources \citep{Leighly2007b}.
Furthermore, the accretion rates in WLQs span the same interval as normal
quasars (Fig.~\ref{fig:mC4all}).

The SEDs of weak line quasars observed in optical/UV band 
(till $\sim$ 1200 \AA) does not differ from the SEDs of normal quasars
\citep[e.g. ][]{DS2009}. However, the far-UV (FUV) spectrum of AGNs 
and their relative quietness in the soft X-ray band produce weak 
emission-lines as supported by photoionization modeling 
\citep[see e.g.][for review]{Leighly2004,LC2007}.
Due to the absent of the far-UV spectra of WLQs we analysed 
the X-ray to optical luminosity ratio \ox\ of different quasars 
(Fig.~\ref{fig:oxC4}).  Similar analysis was carried out by
\citet{Richards2011} or \citet{Wu2011} (see their figure 9 or 
figure 6, respectively). We focus on weak emission-line
quasars and enlarged our sample by adding objects with $\log$ 
EW(\civ) $< 0.6$. Our analysis indicates that the UV/soft X-ray 
SED of WLQs is similar to those of normal AGNs and a soft ionizing 
continuum is not the reason for the weakness of the lines.

\begin{table*}
\caption[]{Arithmetic means and standard deviations of the 
emission-line intensity ratios. All observed line fluxes were dereddened
for the Milky Way contamination.}
\label{tab:ratios}
\centering
\begin{tabular}{l c c c c c c c} 
\hline
\hline
Ratio & \multicolumn{2}{c}{NGC5548} & PG QSO & non-BAL QSO & BAL QSO & 
WLQ(all) & WLQ(sub) \\
(1) & (2a) & (2b) & (3) & (4) & (5) & (6) & (7) \\
\hline
C IV/Ly $\alpha$ & 1.45 &0.94 & $0.46 \pm 0.14$ & 
        &       & $1.62\pm1.57$ & $0.59\pm0.42^{a}$ \\
C IV/Mg II & 
5.80 & 4.98 & $4.38\pm1.54$ & $3.36 \pm1.47$ & $3.03\pm 1.34$& & 
$0.44\pm0.21^{b}$ \\
C III]($\lambda$1909)/Mg II & 
1.07 & 0.91 & $0.99\pm0.20$ & & & & $0.16\pm 0.02^{c}$ \\
Ly $\alpha$/H $\beta$ &
8.57 && $11.75\pm3.25$ &&&& $1.81\pm0.17^{d}$ \\
\hline
\end{tabular}
\\
\begin{flushleft}
Column~(1) refers the names of intensity ratios. In the case of Seyfert 1.5
galaxy NGC~5548 those values are shown in Column~(2a) and (2b). In Column~(2a) 
the \lya, \civ, and C~III] fluxes are taken from 
\citet{Korista95}, the Mg~II flux from \citet{GKN2007}, 
and the \Hb\ flux from \citet{WP96}.
The ratios in Column~(2b) are calculated from corrected for narrow-line 
fluxes and taken from \citet{KG2000}.
Column~(3) refers to sample of 18 radio quiet PG quasars \citep{Shang2007}.
Values of intensity ratio of radio-quiet and radio-intermediate 
97 non-BAL and 14 BAL quasars are shown in Columns~(4) and (5),
respectively. Those sources was selected after cross-matching SDSS~DR7 
Quasar Catalogue \citep{Shen2011} with \citet{Green2009} sample. 
This sample is consistent with sample used in the 
Fig.~\ref{fig:oxC4}. 
In Column~(6) we calculate mean ratio for all WLQs which 
show weak or strong \civ\ lines. Column~(7) refers to 
subsample of WLQs, for which EW(\civ) $< 20$ \AA\ and EW(\lya) $< 15.4$ \AA.
The superscripts in this column correspond to the following information:
$^{a}$ mean is calculated from 59 WLQs, $^{b}$ mean from 
SDSS~J094534 and SDSS~J170109, $^{c}$ value only for SDSS~J094534,
$^{d}$ mean from SDSS~J114153.34+021924.3 and 
SDSS~J123743.08+630144.9. 
Data for intensity of \civ\ line in WLQs are taken from \citet{Shen2011}, 
for \lya\ from \citet{DS2009}, for \civ/Mg~II, 
and C~III]/Mg~II ratios from \citet{Hryn2010}, Hryniewicz et al.
(in preparation), and for H$\beta$ from \citet{Shemmer2010}. 
\end{flushleft}
\end{table*}

The intensity of an emission-line depends on the flux of 
ionizing continuum, \Lion, and on the BLR gas covering factor, 
\covfac: $\mLline \sim \mLion \times \Omega/4\pi$
\citep[see][and his discussion for He II $\lambda$1640]{Ferland2004}. 
The spectral index, \ox, measures by definition the ratio of the 
luminosities at 2 keV and at 2500\AA. If we assume that 
$L_{\nu}(2500 \mathrm{\AA})$ is roughly equal to 
$L_{\nu}(1450 \mathrm{\AA})$ and assuming 
that $L_{\nu}(2 \mathrm{keV}) \simeq$ \Lion, we can write
$\mox \sim \log \mLion - \log L_{\nu}(1450 \mathrm{\AA})$.
We can then express the line equivalent width as:
\begin{displaymath}
\log \mathrm{EW(line)} \approx const_1 + \log \frac{\Omega}{4\pi}
+ \frac{\mox}{const_2}
\end{displaymath}
The correlation EW(\civ)-\ox\ obtained for normal quasars by 
e.g \citet{Wu2009} infers that the gas covering factor in BLR 
in Type 1 quasars is relatively constant. The gas covering factor 
in WLQ objects behaves differently (Fig.~\ref{fig:oxC4}),
suggesting that $\Omega_{\rm WLQ}$~is $\gtrsim 10$ times smaller 
than in QSOs.

Table~\ref{tab:ratios} compares the emission-line intensity ratios 
observed in Seyfert 1.5 galaxy NGC 5548, normal quasars, and WLQs. 
We focus on 59 `real WLQs', i.e. our selected subsample which consists 
of sources with EW(\civ) $\lesssim$ 20 \AA\ and EW(\lya) $<$ 15.4 \AA\   
(see Column~(7) of Table~\ref{tab:ratios}). 
We take into account line intensities
produced by high-ionization lines (HILs; such as \lya, \civ),
intermediate-ionization lines (IILs; e.g. C III]), and 
low-ionization lines (LILs; such as Mg II, H $\beta$).
The \civ/\lya\ intensity ratio for different sources
are the same from a statistical point of view. The ratio of the covering 
factors of the regions responsible for producing \civ\ and \lya\ 
are therefore similar in WLQs and QSOs. Comparing low-, intermediate- 
with high-ionization lines we obtain that the ratios of the covering 
factors of HIL/LIL and IIL/LIL are lower in WLQs that in normal quasars. 
Even if for WLQ these ratios are based on only few sources. This suggests 
that the covering factor of the BLR is smaller in WLQ.

This is in agreement with observations of the weak H$\beta$ 
emission-lines in SDSS J114153.34+021924.3 and SDSS J123743.08+630144.9
\citep{Shemmer2010}. There authors have explained the weakness of their 
emission-lines by a deficit of the BLR.
The absence of BLR in WLQs have also been recently suggested 
by \citet{LZ2011}. The existence of bright AGNs with dusty tori, 
but without BLR could be understood when an anisotropic radiative pressure 
is released from an accretion disc. \citeauthor{LZ2011} stated 
that this is possible just before the normal 
phase of an AGN. Additionally, \citet{LM2004} suggested 
based on observations of the emission-line profiles of NLS1 galaxies 
IRAS~13224-3809 and 1H~0707-495
that the high-ionization lines are produced in a wind
and that the intermediate- and low-ionization lines are produced 
in low-velocity gas associated with the accretion disk at the base of 
the wind. Both pictures are consistent with a suggestion that 
the regions producing emission-lines are developed by winds 
\citep{Hawkins2004,Hryn2010}. In that case, when the BLR is created 
its covering factor is lower than estimated in normal AGNs.

There is an observational analogy between weak emission-line quasars
and the class of ``optical dull'' AGNs (also called XBONGs -- 
``X-ray bright, optically normal galaxies''). Their X-ray emission 
is bright while they lack both the broad emission-lines of Type~1 AGNs 
and the narrow emission-lines Type~2 AGNs 
\citep{Elvis81,Comastri2002, GeorganGeorgak2005}.
There are a few hypothesis trying to answer the latent nature of XBONGs
\citep[see e.g.][]{Moran2002,Severgnini2003,Rigby2006,Civano2007,Trump2009}.
However, none of them (such as dilution their spectra by a host galaxy,
the low Eddington accretion rate) can explain WLQs.

\citet{Elvis2000} has proposed an empirically derived structure for
quasars. He suggests presence of the funnel-shaped geometrically thin 
accretion outflow which contains an high ionized gas embedded
in the colder BLR clouds.  According to our paper the low covering
factor of the BLR means that WLQ has got less clouds in the
outflow or equivalently the ``funnel'' wind is geometrically
thiner.

Low covering factor of the BLR in WLQs would have additional consequence 
observed in the infrared (IR) band. \citet{GKN2007} have argued that 
the covering factors of the BLR and of the dusty torus have to be the 
same. It means that a small BLR in WLQs causes an evaporation of dust 
in the torus and a reduction of its IR emissivity. \citet{DS2009} 
mentioned that two weak line quasars SDSS~J140850.91+020522.7 
(with EW(\civ) = 1.95 \AA) and SDSS~J144231.72+011055.2 
(with EW(\civ) = 16.9 \AA) are fainter in the IR ($\sim 24 \mu$m) band
by 30-40\%. Additionally, the IR flux density of SDSS~J130216.13+003032.1 
(EW(\civ) = 27.8 \AA) is also relatively low. More IR observations 
of WLQs are required to confirm this hypothesis.

\section{Conclusions}
\label{sec:con}

We have explored the Baldwin effect (BEff) in 
82 high-redshift (z $>$ 2.2) and 2 intermediate-redshift
weak-line quasars (WLQs) and compared them  with a set of normal 
quasars.
We draw the following conclusions:
\begin{itemize}
\item The relationship between the rest-frame equivalent width for 
\civ\ emission-line and the Eddington ratio
observed in WLQs has different normalization than for normal QSOs. 
This shift disagrees with the super-Eddington
hypothesis \citep[e.g.][]{Shemmer2010}.

\item The weakness or even the absence of emission-lines in WLQs is likely
caused by a low covering factor of the broad line region (BLR) rather 
than by a very soft ionizing continuum. The comparison of the EW(\civ) 
and of the spectral indices, \ox, shows that the gas covering factor of 
the BLR in WLQs is $\gtrsim 10$ times less than for normal QSOs.

\item The ratios of the covering factors of regions responsible 
for producing \civ\ and \lya\ are similar in WLQs and QSOs.

\item The ratios of the covering factors $\Omega_{\rm HIL}/\Omega_{\rm LIL}$ 
are lower in WLQ than in QSOs showing the deficit of the BLR in WLQ.
However, this result is based on observations of only four sources. 

\item The radio-intermediate quasar SDSS~J170108.89+395443.0 (z $=1.89$) 
is a new intermediate-redshifted WLQ with rest-frame 
EW(\civ) $= 2.09$\AA\ and EW(Mg II) $= 9.41$\AA, respectively 
\citep{Shen2011}.

\item The definition of WLQ objects should take into account not only 
the weakness of \lya\ or \civ\ emission-lines, separately, but both lines
together. `False WLQs' (sources with prominent \civ) are probably normal 
Type~1 quasars with intervening \lya\ absorption.
\end{itemize}

\section*{Acknowledgments}

We would like to thank an anonymous referee for useful comments 
that improve our paper. We are grateful to Bozena Czerny, Krzysztof Hryniewicz
and Joanna Kuraszkiewicz for advices during calculation and doing our analysis.
We also thank Gary Ferland for pointing out a helpful article.
MN thanks the Scientific Exchange Programme (Sciex) NMS$^{\rm ch}$
for opportunity of working at the ISDC.
This research has been supported in part by the Polish MNiSW grants
NN203 380136, and 362/1/N-INTEGRAL/2008/09/0.

\bibliographystyle{mn2e}
\bibliography{BEffCovFactor}


\onecolumn
%
%


\begin{longtable}{lccccccr}
\caption{The sample of weak emission-line quasars. 
\label{tab:WLQ}} \\
\hline\hline
Name & $z_{\rm SDSS}$ & EW(\civ) & ref. & $\log M_{\rm BH}$ & \ox & ref. & $\log \mLL$ \\
     &              & (\AA)    & & ($\mathrm{M_{\odot}}$) &     &      &     \\
\endfirsthead
\caption{continued.} \\
\hline\hline
Name & $z_{\rm SDSS}$ & EW(\civ) & ref. & $\log M_{\rm BH}$ & \ox & ref. & $\log \mLL$ \\
    &              & (\AA)    & & ($\mathrm{M_{\odot}}$) &     &      &     \\
\hline\hline
\endhead
\hline\hline
\hline
SDSS~010802.90-010946.1 & 3.330 & $\le 15.1$ & & $10.243 \pm 0.390$ & & & $-1.308 \pm 0.390$ \\
SDSS~025646.56+003858.3 & 3.473 & $ 6.3 \pm 1.2 $ & (1) & $ 9.968 \pm 0.532$ & & & $-0.865 \pm 0.532$ \\
SDSS~031712.23-075850.3 & 2.695 & $\le 1.0$ & (2)& & $-1.58$ & (3) & \\
SDSS~080523.32+214921.1 & 3.463 & $23.4 \pm 3.0$ & & $ 9.107 \pm 0.662$ & & & $-0.042 \pm 0.662$ \\
SDSS~080906.87+172955.1 & 2.953 & $ 3.6 \pm 0.7$ & & $ 9.509 \pm 0.489$ & & & $-0.185 \pm 0.489$ \\
SDSS~082059.34+561021.9 & 3.636 & $31.6 \pm 4.7$ & & $ 9.376 \pm 0.089$ & & & $-0.457 \pm 0.090$ \\
SDSS~082638.59+515233.2 & 2.850 & $ 4.3 \pm 0.5$ & & $10.369 \pm 0.050$ & & & $-0.610 \pm 0.050$ \\
SDSS~083122.57+404623.3 & 4.885 & $\le 10.5$     & & $ 9.727 \pm 0.621$ & & & $-0.311 \pm 0.621$ \\
SDSS~083304.73+415331.3 & 2.329 & $\le 4.2$      & & $ 9.107 \pm 0.471$ & & & $-0.467 \pm 0.471$ \\
SDSS~083330.56+233909.1 & 2.417 & $\le 5.4$      & & $ 9.199 \pm 0.464$ & & & $-0.499 \pm 0.465$ \\
SDSS~084249.03+235204.7 & 3.316 & $\le 9.7$      & & $ 9.422 \pm 0.247$ & & & $-0.512 \pm 0.247$ \\
SDSS~084424.24+124546.5 & 2.505 & $\le 3.5$      & & $ 9.592 \pm 0.197$ & & & $-0.315 \pm 0.197$ \\
SDSS~084434.15+224305.2 & 3.117 & $42.1 \pm 5.6$ & & $ 9.860 \pm 0.229$ & & & $-1.016 \pm 0.229$ \\
SDSS~090703.91+410748.3 & 2.672 & $ \le 1.9$     & & $ 8.727 \pm 0.101$ & & & $ 0.129 \pm 0.102$ \\
SDSS~091738.90+082053.9 & 3.252 & $18.0 \pm 0.8$ & (1) & $ 9.853 \pm 0.124$ & & & $-0.948 \pm 0.125$ \\
SDSS~092312.75+174452.8 & 2.260 & $\le 7.5$      & & $ 9.960 \pm 0.414$ & & & $-1.117 \pm 0.414$ \\
SDSS~093306.88+332556.6 & 4.560 & $\le 37.3$     & & $10.073 \pm 0.744$ & & & $-0.946 \pm 0.744$ \\
SDSS~094533.98+100950.1 & 1.661 & $\le 4.0$      & & $ 9.786 \pm 0.036$ & & & $-0.753 \pm 0.036$ \\
SDSS~095108.76+314705.8 & 3.032 & $\le 12.1$     & & $ 9.770 \pm 0.507$ & & & $-1.237 \pm 0.508$ \\
SDSS~101204.04+531331.7 & 2.990 & $\le 3.7$      & & $ 9.510 \pm 0.398$ & $-1.49$ & (3) & $-0.736 \pm 0.399$ \\
SDSS~101849.78+271914.9 & 2.603 & $\le 4.9$      & & $ 8.331 \pm 0.956$ & & & $ 0.322 \pm 0.956$ \\
SDSS~102609.92+253651.2 & 2.317 & $\le 3.3$      & & $ 8.998 \pm 0.515$ & & & $-0.203 \pm 0.515$ \\
SDSS~102949.80+605731.7 & 3.190 & $37.1 \pm 8.6$ & & $ 9.624 \pm 0.128$ & & & $-1.036 \pm 0.131$ \\
SDSS~103240.54+501211.0 & 3.874 & $\le 12.5$     & & $ 9.779 \pm 0.140$ & & & $-0.531 \pm 0.142$ \\
SDSS~104650.29+295206.8 & 4.266 & $\le 37.8$     & & $10.530 \pm 0.317$ & & & $-1.599 \pm 0.321$ \\
SDSS~105049.27+441144.7 & 4.320 & $\le 9.7$      & & $10.039 \pm 0.573$ & & & $-0.808 \pm 0.574$ \\
SDSS~111642.81+420324.9 & 2.526 & $\le 3.2$      & & $ 9.142 \pm 1.604$ & & & $-0.198 \pm 1.604$ \\
SDSS~113354.89+022420.9 & 3.990 & $16.1 \pm 2.1$ & & $10.014 \pm 0.092$ & & & $-0.782 \pm 0.093$ \\
SDSS~113729.42+375224.2 & 4.166 & $\le 7.4$      & & $ 8.193 \pm 1.069$ & & & $ 0.740 \pm 1.069$ \\
SDSS~113747.64+391941.5 & 2.395 & $\le 5.9$      & & $ 8.507 \pm 1.167$ & & & $ 0.224 \pm 1.167$ \\
SDSS~114153.33+021924.3 & 3.480 & $\le 1.0$ & (2)& & $-1.54$& (3) & \\
SDSS~114412.76+315800.8 & 3.235 & $ 6.8 \pm 1.0$ & & $10.128 \pm 0.143$ & & & $-0.884 \pm 0.143$ \\
SDSS~114958.53+375115.0 & 4.315 & $\le 41.5 $    & & $ 8.310 \pm 1.875$ & & & $ 0.527 \pm 1.876$ \\
SDSS~115254.96+150707.7 & 3.328 & $10.4 \pm 1.2$ & (1) & $10.080 \pm 0.450$ & & & $-0.807 \pm 0.450$ \\
SDSS~115308.45+374232.1 & 3.028 & $17.2 \pm 1.1$ & (1) & $ 9.690 \pm 0.211$ & & & $-0.894 \pm 0.212$ \\
SDSS~115906.52+133737.7 & 3.984 & $13.0 \pm 1.2$ & & $10.664 \pm 0.055$ & & & $-0.889 \pm 0.055$ \\
SDSS~115933.53+054141.6 & 3.286 & $29.1 \pm 1.6$ & (1) & $ 9.425 \pm 0.262$ & & & $-0.791 \pm 0.263$ \\
SDSS~115959.71+410152.9 & 2.788 & $ 8.7 \pm 1.1$ & & $10.427 \pm 0.063$ & & & $-0.895 \pm 0.063$ \\
SDSS~120059.68+400913.1 & 3.366 & $72.6 \pm 0.5$ & & $ 8.177 \pm 0.775$ & & & $ 0.645 \pm 0.776$ \\
SDSS~121221.56+534127.8 & 3.097 & $\le 3.0$      & & & $-1.93$ & (3) & \\
SDSS~121812.39+444544.5 & 4.518 & $\le 20.4$     & & $10.314 \pm 0.621$ & & & $-1.216 \pm 0.622$ \\
SDSS~122021.39+092135.8 & 4.110 & $25.2 \pm 1.3$ & (1) & $ 8.971 \pm 1.429$ & & & $ 0.264 \pm 1.430$ \\
SDSS~122359.35+112800.0 & 4.120 & $\le 37.6$     & & $ 9.781 \pm 0.161$ & & & $-0.624 \pm 0.163$ \\
SDSS~122445.26+375921.3 & 4.315 & $\le 18.9$     & & $ 9.314 \pm 1.205$ & & & $-0.414 \pm 1.207$ \\
SDSS~123116.08+411337.3 & 3.838 & $\le 12.1$     & & $ 9.732 \pm 0.458$ & & & $-0.816 \pm 0.458$ \\
SDSS~123132.37+013814.0 & 3.229 & $\le 1.9$      & & $ 9.997 \pm 0.670$ & $-1.37$ & (3) & $-0.742 \pm 0.670$ \\
SDSS~123315.94+313218.4 & 3.222 & $ 6.4 \pm 1.4$ & (1) & $ 9.163 \pm 1.131$ & & & $-0.469 \pm 1.131$ \\
SDSS~123540.19+123620.7 & 3.215 & $51.2 \pm 9.5$ & (1) & $ 9.370 \pm 0.183$ & & & $-0.612 \pm 0.183$ \\
SDSS~123743.08+630144.8 & 3.425 & $ 7.7 \pm 1.1$ & (1) & $ 8.983 \pm 0.901$ & $< -1.55$ & (3) & $ 0.119 \pm 0.901$ \\
SDSS~124204.28+625712.1 & 3.321 & $ 8.6 \pm 1.5$ & (1) & $ 8.708 \pm 1.029$ & & & $ 0.079 \pm 1.029$ \\
SDSS~124745.39+325147.0 & 2.249 & $\le 4.4$      & & $ 9.208 \pm 0.363$ & & & $-0.444 \pm 0.364$ \\
SDSS~125306.73+130604.9 & 3.624 & $10.8 \pm 1.5$ & & $ 9.949 \pm 0.112$ & & & $-0.670 \pm 0.113$ \\
SDSS~125319.10+454152.8 & 3.435 & $\le 14.0$     & & $10.139 \pm 0.492$ & & & $-1.397 \pm 0.493$ \\
SDSS~130216.13+003032.1 & 4.506 & $\le 1.0$ & (2) & & $-2.08$ & (3) & \\
SDSS~131429.00+494149.0 & 3.813 & $13.9 \pm 1.1$ & (1) & $10.221 \pm 0.134$ & & & $-1.040 \pm 0.135$ \\
SDSS~132603.00+295758.1 & 3.767 & $ 6.0 \pm 0.5$ & (1) & $ 9.802 \pm 0.309$ & & & $-0.434 \pm 0.309$ \\
SDSS~132703.26+341321.7 & 2.558 & $ 6.6 \pm 1.3$ & & $ 9.512 \pm 0.562$ & & & $-0.609 \pm 0.562$ \\
SDSS~133146.20+483826.5 & 3.742 & $20.5 \pm 3.6$ & & $ 9.949 \pm 0.142$ & & & $-0.803 \pm 0.142$ \\
SDSS~133422.63+475033.5 & 4.950 & $\le 6.9$      & (2) & & $-1.70$ & (3) & \\
SDSS~134521.39+281822.2 & 4.082 & $\le 17.8$     & & $10.141 \pm 0.700$ & & & $-1.290 \pm 0.701$ \\
SDSS~140300.23+432805.4 & 4.696 & $\le 25.9$     & & $ 9.856 \pm 0.943$ & & & $-0.768 \pm 0.943$ \\
SDSS~140850.91+020522.7 & 4.007 & $\le 3.2$      & & & $-1.54$ & (4) & \\
SDSS~141209.96+062406.9 & 4.466 & $\le 12.7$     & & $ 9.859 \pm 1.217$ & & & $-0.708 \pm 1.218$ \\
SDSS~141318.86+450523.0 & 3.113 & $45.0 \pm 5.4$ & & $ 9.095 \pm 0.095$ & & & $-0.202 \pm 0.096$ \\
SDSS~141657.93+123431.6 & 2.603 & $\le 3.9$      & & $ 9.159 \pm 0.793$ & & & $-0.126 \pm 0.793$ \\
SDSS~142103.83+343332.0 & 4.907 & $\le 50.5$     & & $ 9.841 \pm 0.480$ & $-2.07$ & (3) & $-0.362 \pm 0.481$ \\
SDSS~142144.98+351315.4 & 4.556 & $39.8 \pm 5.9$ & & $ 9.884 \pm 0.351$ & & & $-0.531 \pm 0.351$ \\
SDSS~142257.67+375807.4 & 3.163 & $39.5 \pm 1.8$ & (1) & $ 9.293 \pm 0.609$ & & & $-0.560 \pm 0.610$ \\
SDSS~144127.65+475048.7 & 3.190 & $24.9 \pm 1.9$ & & $ 8.598 \pm 1.069$ & & & $ 0.466 \pm 1.069$ \\
SDSS~144231.71+011055.3 & 4.507 & $\le 22.5$     & & & $-1.42$ & (4) & \\
SDSS~144803.36+240704.2 & 3.544 & $\le 12.5$     & & $ 9.967 \pm 0.209$ & & & $-0.905 \pm 0.210$ \\
SDSS~155645.30+380752.8 & 3.320 & $\le 2.4$      & & $ 8.407 \pm 1.114$ & & & $ 0.536 \pm 1.115$ \\
SDSS~160336.64+350824.3 & 4.460 & $12.9 \pm 2.3$ & & $10.296 \pm 0.102$ & & & $-0.820 \pm 0.102$ \\
SDSS~161122.45+414409.5 & 3.131 & $19.6 \pm 2.4$ & & $ 9.940 \pm 0.165$ & & & $-0.829 \pm 0.166$ \\
SDSS~163411.82+215325.0 & 4.529 & $\le 29.3$     & & $10.681 \pm 0.221$ & & & $-1.248 \pm 0.221$ \\
SDSS~170108.89+395443.0 & 1.889 & $\le 3.9$      & & $ 8.516 \pm 0.568$ & $-1.27$ & (5) & $-0.056 \pm 0.568$ \\
SDSS~210216.52+104906.6 & 4.182 & $16.6 \pm 1.0$ & (1) & $10.126 \pm 0.440$ & & & $-0.817 \pm 0.441$ \\
SDSS~214753.29-073031.3 & 3.153 & $ 6.5 \pm 0.8$ & (1) & $ 9.786 \pm 0.638$ & & & $-0.868 \pm 0.638$ \\
SDSS~223827.17+135432.6 & 3.516 & $28.5 \pm 3.4$ & & $ 9.868 \pm 0.121$ & & & $-0.923 \pm 0.122$ \\
SDSS~225246.43+142525.8 & 4.904 & $\le 18.6$     & & $ 9.786 \pm 0.340$ & & & $-0.704 \pm 0.341$ \\
SDSS~233255.72+141916.3 & 4.754 & $\le 45.1$     & & $ 8.665 \pm 1.053$ & & & $ 0.009 \pm 1.055$ \\
SDSS~233446.40-090812.2 & 3.317 & $10.4 \pm 1.2$ & & $10.143 \pm 0.110$ & & & $-0.571 \pm 0.110$ \\
SDSS~233939.48-103539.3 & 2.757 & $\le 5.2$      & & $ 9.837 \pm 0.525$ & & & $-0.719 \pm 0.525$ \\
\hline\hline
\end{longtable}
\noindent
Column~(1) lists the object name. Column~(2) shows the redshift taken from  
the SDSS DR7 Quasar Catalogue \citep{Shen2011}. Columns~(3),(5), and (8) list
the rest frame equivalent width of \civ\ emission-line,  logarithm of black hole mass, and 
accretion rates in the Eddington units, respectively. Column~(6) refers to the X-ray to optical luminosity ratio.
The values without reference were taken from the SDSS DR7 Quasar Catalogue. 
A few EW(\civ) and all \ox\ values were taken from articles with references shown in Columns~(4) and (7).
Numbers in parentheses correspond to the following references: 
(1) -- \citet{DS2009}, (2) -- this paper,
(3) -- \citet{Shemmer2009}, (4) -- \citet{Shemmer2006}, (5) -- \citet{Green2009}.

\label{lastpage}
\end{document}